\documentclass[twoside]{article}
\usepackage{qic}

\usepackage{amsmath}
\usepackage{amssymb}
\usepackage{braket}

\newtheorem{thm}{Theorem}
\newtheorem{cor}{Corollary}
\newtheorem{lem}{Lemma}

\def\erf{\qopname\relax{no}{erf}}

\begin{document}
\setlength{\textheight}{8.0truein}    

\runninghead{Realization of the probability laws in the quantum central limit theorems by a quantum walk}
            {T. Machida}

\normalsize\textlineskip
\thispagestyle{empty}
\setcounter{page}{1}

\vspace*{0.88truein}

\alphfootnote

\fpage{1}

\centerline{\bf
Realization of the probability laws in the quantum central limit theorems by a quantum walk}
\vspace*{0.37truein}
\centerline{\footnotesize
Takuya Machida}
\vspace*{0.015truein}
\centerline{\footnotesize\it Meiji Institute for Advanced Study of Mathematical Sciences,}
\baselineskip=10pt
\centerline{\footnotesize\it Meiji University, 1-1-1 Higashimita, Tamaku, Kawasaki 214-8571, Japan}
\vspace*{0.225truein}

\vspace*{0.21truein}

\abstracts{
Since a limit distribution of a discrete-time quantum walk on the line was derived in 2002, a lot of limit theorems for quantum walks with a localized initial state have been reported.
On the other hand, in quantum probability theory, there are four notions of independence (free, monotone, commuting, and boolean independence) and quantum central limit theorems associated to each independence have been investigated.
The relation between quantum walks and quantum probability theory is still unknown.
As random walks are fundamental models in the Kolmogorov probability theory, can the quantum walks play an important role in quantum probability theory?
To discuss this problem, we focus on a discrete-time 2-state quantum walk with a non-localized initial state and present a limit theorem.
By using our limit theorem, we generate probability laws in the quantum central limit theorems from the quantum walk.
}{}{}

\vspace*{10pt}

\vspace*{3pt}

\vspace*{1pt}\textlineskip    

\bibliographystyle{qic}

\section{Introduction}

Quantum walks (QWs), which are quantum versions of random walks, are expected to be one of the simple dynamics to understand quantum systems, and they are permeating more and more in science.
The behavior of the QWs is different from that of random walks.
With a proper rescaling, probability distributions of the QWs on the line $\mathbb{Z}=\left\{0,\pm 1,\pm2,\ldots\right\}$ after many steps are approximately expressed by probability density functions with a compact support.
Even as time evolution of the QWs is determined by a space-homogeneous dynamics, limit density functions of the QWs have singularity in space. 
To interpret the interesting property of the QWs, a lot of long-time limit theorems have been investigated, because getting such a limit theorem is equivalent to understanding asymptotic behavior of the QWs after long time.
To forecast the behavior of the QWs, it is worth studying the long-time limit theorems.

On the other hand, quantum probability theory is algebraic generalization of the Kolmogorov probability theory.
One of the major ideas in probability is the notion of independence.
In quantum probability theory, there are four notions of independence (free, monotone, commuting, and boolean independence) and quantum central limit theorems associated to each independence have been proved~\cite{Voiculescu1985,Lu1997,Muraki1997,CushenHudson1971,GiriWaldenfels1978,SpeicherWoroudi1997}.
As we well know, a long-time limit distribution of simple random walks, that is the Gaussian distribution, follows from the central limit theorem for independence in the Kolmogorov probability theory.
The relation between the random walks and the central limit theorem has supported fruitful applications of the random walks.
Meanwhile, there are only a few current applications of the QWs, which are mainly construction of quantum algorithms in quantum computation (see, for instance, \cite{Ambainis2003,ShenviKempeWhaley2003,ChildsCleveJordanYonge-Mallo2009,FeldmanHilleryLeeReitznerZhengBuvzek2010,ReitznerHilleryFeldmanBuvzek2009,Venegas-Andraca2008}).
If the QWs connect to the quantum central limit theorems, they would be more useful via quantum probability theory.
To discuss the potential of the QWs in quantum probability theory, we focus on the limit distribution of a discrete-time QW on the line in this paper.

The first derivation of a limit theorem for discrete-time 2-state QWs on the line was done by a path counting method in 2002~\cite{Konno2002} (see also Konno~\cite{Konno2005}). 
In probability distributions of the QWs on time- or space-inhomogeneous environments, localization can occur~\cite{Konno2010,KonnoLuczakSegawa2013,Machida2011}.
Inui et al.~\cite{InuiKonnoSegawa2005} obtained a limit distribution with localization for a 3-state QW.
In addition, localization of multi-state walks was also discussed in Inui and Konno~\cite{InuiKonno2005} and Segawa and Konno~\cite{SegawaKonno2008}.
Konno and Machida~\cite{KonnoMachida2010} rewrote a 2-state QW with memory introduced by McGettrick~\cite{McGettrick2010} as a 4-state QW and got two limit theorems.
Konno~\cite{Konno2009} calculated the QW on random environments whose probability distribution after long time isn't localized at all.
Via a weak convergence theorem, Chisaki et al.~\cite{ChisakiKonnoSegawaShikano2011} investigated crossover from QWs to random walks.
Recently, Konno et al.~\cite{KonnoMachidaWakasainpress} found relations between long-time limit density functions of discrete- or continuous-time QWs on the line and well-known Fucksian linear differential equations of the second order (the Heun equation, the Gauss hypergeometric equation).  
These limit theorems are results for the QWs with a localized initial condition.
Since we have not obtained the probability density functions related with the central limit theorems in quantum probability theory from the QWs with a localized initial state, we will move our focus to the QWs with a non-localized initial state in this paper.
There are a few results for the QWs distributed widely in an initial state~\cite{AbalDonangeloRomanelliSiri2006,AbalSiriRomanelliDonangelo2006,ValcarcelRoldanRomanelli2010,ChandrashekarBusch2012}.
Abal et. al.~\cite{AbalDonangeloRomanelliSiri2006,AbalSiriRomanelliDonangelo2006} analyzed a QW starting from two positions.
Chandrashekar and Busch~\cite{ChandrashekarBusch2012} reported numerical results for a QW with an initial state ranging over an area.
Valc{\'a}rcel et al.~\cite{ValcarcelRoldanRomanelli2010} treated the QW initialized by a Gaussian-like distribution in a continuum limit.
A uniform stationary measure of the Hadamard walk with a non-localized initial state was discussed in Konno et al.~\cite{KonnoLuczakSegawa2013}.

In the rest of this paper, we deal with the following topics.
We introduce the definition of a discrete-time 2-state QW on the line in Sec.~\ref{sec:definition}.
Section~\ref{sec:limit_th} is devoted to show our limit theorem.
In addition, the limit theorem derives a corollary. 
After introducing a non-localized initial state which is defined in Sec.~\ref{sec:initial}, we report that the QW generates the probability laws associated to the independence in quantum probability theory.
In the final section, we conclude our results and discuss a future problem.

\section{Definition of a discrete-time QW on the line}
\label{sec:definition}
Total system of discrete-time 2-state QWs on the line is defined in a tensor space $\mathcal{H}_p\otimes\mathcal{H}_c$, where $\mathcal{H}_p$ is called a position Hilbert space which is spanned by a basis $\left\{\ket{x}:\,x\in\mathbb{Z}\right\}$ and $\mathcal{H}_c$ is called a coin Hilbert space which is spanned by a basis $\left\{\ket{0},\ket{1}\right\}$ with the vectors $\bra{0}=[1,0],\,\bra{1}=[0,1]$.
Let $\ket{\psi_{t}(x)} \in \mathcal{H}_c$ be the probability amplitudes of the walker at position $x$ at time $t \in\left\{0,1,2,\ldots\right\}$.
The state of the 2-state QWs on the line at time $t$ is expressed by $\ket{\Psi_t}=\sum_{x\in\mathbb{Z}}\ket{x}\otimes\ket{\psi_{t}(x)}$.
Time evolution of the QWs is described by a unitary matrix
\begin{equation}
 U=\cos\theta\ket{0}\bra{0}+\sin\theta\ket{0}\bra{1}+\sin\theta\ket{1}\bra{0}-\cos\theta\ket{1}\bra{1}
\end{equation}
with $\theta\in [0,2\pi)$.
The behavior of the QW with $\theta= 0, \pi/2, \pi, 3\pi/2$ is trivial.
So, in the present paper we don't treat such a case.
The amplitudes evolve according to
\begin{equation}
 \ket{\psi_{t+1}(x)}=\ket{0}\bra{0}U\ket{\psi_t(x+1)}+\ket{1}\bra{1}U\ket{\psi_t(x-1)}.
\end{equation}
The probability that the quantum walker $X_t$ can be found at position $x$ at time $t$ is defined by
\begin{equation}
 \mathbb{P}(X_t=x)=\braket{\psi_t(x)|\psi_t(x)}. 
\end{equation}
By giving the initial states $\ket{\psi_0(x)}$, we can determine the probability distribution $\mathbb{P}(X_t=x)$ for any time $t$.
In this paper, we focus on a special non-localized initial state which will be introduced in Sec.~\ref{sec:initial}.

\section{Limit theorem}
\label{sec:limit_th}
In this section, we present a limit theorem of the QW as $t\to\infty$ by using the Fourier analysis which is one of the standard methods to derive limit theorems of the QWs~\cite{Konno2010,Machida2011,InuiKonnoSegawa2005,InuiKonno2005,SegawaKonno2008,KonnoMachida2010,ChisakiKonnoSegawaShikano2011,GrimmettJansonScudo2004}.
The long-time limit theorem plays an essential role to draw asymptotic behavior of the QW after many steps.
The time evolution of the QW provides the Fourier transform $\ket{\hat{\Psi}_{t}(k)}=\sum_{x\in\mathbb{Z}} e^{-ikx}\ket{\psi_t(x)}\,(k\in\left[-\pi,\pi\right))$, the relation $\ket{\hat{\Psi}_{t+1}(k)}= \hat U(k)\ket{\hat{\Psi}_{t}(k)}=\hat U(k)^t \ket{\hat\Psi_{0}(k)}$ , where $\hat U(k)=(e^{ik}\ket{0}\bra{0}+e^{-ik}\ket{1}\bra{1})U$.
After straightforward calculation by the Fourier analysis, for $r=0,1,2,\ldots$, we get
\begin{align}
 \lim_{t\to\infty}\mathbb{E}\biggl[\biggl(\frac{X_t}{t}\biggr)^r\biggr]=&\int_{0}^{\pi} \frac{dk}{2\pi} h(k)^r\,\biggl[\biggl\{\left|\braket{v(k)|\hat\Psi_0(k)}\right|^2+\left|\braket{v(-k)|\hat\Psi_0(-k)}\right|^2\biggr\}\nonumber\\
 &+(-1)^r\biggl\{\left|\braket{\overline{v(\pi-k)}|\hat\Psi_0(k)}\right|^2+\left|\braket{\overline{v(\pi+k)}|\hat\Psi_0(-k)}\right|^2\biggr\}\biggr],
\end{align}
where
\begin{align}
 h(k)=&\frac{c\cos k}{\sqrt{1-c^2\sin^2 k}},\\
 \ket{v(k)}=&\frac{e^{ik}s}{\sqrt{N(k)}}\ket{0}-\frac{c\cos k+\sqrt{1-c^2\sin^2 k}}{\sqrt{N(k)}}\ket{1},\\
 N(k)=&1+s^2+c^2\cos 2k+2c\cos k\sqrt{1-c^2\sin^2 k},
\end{align}
and $c=\cos\theta, s=\sin\theta$.
The detail can be found in Grimmett et al.~\cite{GrimmettJansonScudo2004}.
By putting $h(k)=x$, we obtain the following lemma.

\begin{lem}
For $r=0,1,2,\ldots$, we have
\begin{equation}
 \lim_{t\to\infty}\mathbb{E}\left[\left(\frac{X_t}{t}\right)^r\right]=\int_{-\infty}^\infty x^r \frac{|s|}{2\pi(1-x^2)\sqrt{c^2-x^2}}\eta(x)I_{(-|c|,|c|)}(x)\,dx,
\end{equation}
where $I_A(x)=1$ if $x\in A$, $I_A(x)=0$ if $x\notin A$ and
 \begin{align}
  \eta(x)=&\left|\braket{v(\kappa(x))|\hat\Psi_0(\kappa(x))}\right|^2+\left|\braket{v(\kappa(x))|\hat\Psi_0(\kappa(x)-\pi)}\right|^2\nonumber\\
  &+\left|\braket{v(-\kappa(x))|\hat\Psi_0(-\kappa(x))}\right|^2+\left|\braket{v(-\kappa(x))|\hat\Psi_0(\pi-\kappa(x))}\right|^2,\\
  \kappa(x)=&\arccos\left(\frac{|s|x}{c\sqrt{1-x^2}}\right)\,\in\left[0,\pi\right].
 \end{align}
\label{lem:limit}
\end{lem}

If we give a special initial state to the Fourier transform $\ket{\hat\Psi_0(k)}$, the following limit theorem, which can be immediately proved from Lemma~\ref{lem:limit}, is obtained.

\begin{thm}
Let $F:\mathbb{R}\longrightarrow\mathbb{R}$ be a function that satisfies $F(k+2\pi)=F(k)$, $\int_{-\pi}^{\pi}F(k)^2 dk=2\pi$ and $F(k)\in C^{\infty}[-\pi,\pi]$ almost everywhere, where $\mathbb{R}$ means the set of real numbers.
If we assume $\ket{\hat\Psi_0(k)}=F(k)(\alpha\ket{0}+\beta\ket{1})$ with $\alpha,\beta\in\mathbb{C}$ and $|\alpha|^2+|\beta|^2=1$,
\begin{equation}
 \lim_{t\to\infty}\mathbb{E}\biggl[\left(\frac{X_t}{t}\right)^r\biggr]=\int_{-\infty}^\infty x^r\, \Bigl\{f_1(x;\alpha,\beta)\eta_1(x)+f_2(x;\alpha,\beta)\eta_2(x)\Bigr\}I_{(-|c|,|c|)}(x)\,dx,
\end{equation}
where $\mathbb{C}$ is the set of complex numbers,
\begin{align}
 f_1(x;\alpha,\beta)=&\frac{|s|}{\pi(1-x^2)\sqrt{c^2-x^2}}\biggl[1-\biggl\{|\alpha|^2-|\beta|^2+\frac{2s\Re(\alpha\overline{\beta})}{c}\biggr\}x\biggr],\\
 f_2(x;\alpha,\beta)=&-\frac{s\Im(\alpha\overline{\beta})}{|c|\pi(1-x^2)},\\
 \eta_1(x)=&\frac{1}{4}\Bigl\{F(\kappa(x))^2+F(-\kappa(x))^2+F(\kappa(x)-\pi)^2+F(\pi-\kappa(x))^2\Bigr\},\\
 \eta_2(x)=&\frac{1}{2}\Bigl\{F(\kappa(x))^2-F(-\kappa(x))^2+F(\kappa(x)-\pi)^2-F(\pi-\kappa(x))^2\Bigr\},
\end{align}
\label{th:limit}
\end{thm}
and $\Re(z)$ (resp. $\Im(z)$) denotes the real (resp. imaginary) part of $z\in\mathbb{C}$.
Theorem~\ref{th:limit}, moreover, leads us to the following corollary.

\begin{cor}
If the function $F(k)$ satisfies $|F(k-\pi)|=|F(-k)|=|F(k)|$,
\begin{equation}
 \lim_{t\to\infty}\mathbb{E}\bigg[\bigg(\frac{X_t}{t}\biggr)^r\biggr]=\int_{-\infty}^\infty  x^r f_1(x;\alpha,\beta)F(\kappa(x))^2 I_{(-|c|,|c|)}(x)\,dx.
\end{equation}
\label{cor:1}
\end{cor}
In Sec.~\ref{sec:example}, we will show four examples of the QW as $t\to\infty$ by using our limit theorem.

\section{Non-localized initial state}
\label{sec:initial}
In this section, by using the Fourier series expansion, we construct a non-localized initial state. 
Let $w:\mathbb{R}\longrightarrow\mathbb{R}$ be the function that satisfies $w(k+2\pi)=w(k)$ and $w\in L^2[-\pi,\pi]$.
Assuming $W(w)=\int_{-\pi}^{\pi} w(k)^2\,dk > 0$, we take the initial state of 2-state QWs on the line as
\begin{equation}
 \ket{\psi_0(x)}=\frac{1}{\sqrt{2\pi W(w)}}\left(\int_{-\pi}^{\pi} w(k)e^{ikx}dk\right)\left(\alpha\ket{0}+\beta\ket{1}\right),\label{eq:upis}
\end{equation}
with $\alpha,\beta\in\mathbb{C}$ and $|\alpha|^2+|\beta|^2=1$.
We should note that $\sum_{x\in\mathbb{Z}}\mathbb{P}(X_0=x)=\sum_{x\in\mathbb{Z}}\braket{\psi_0(x)|\psi_0(x)}=1$ by Parseval's theorem.
The Carleson-Hunt theorem changes the initial state of the Fourier transform,
\begin{align}
 \ket{\hat\Psi_0(k)}=\sqrt{\frac{2\pi}{W(w)}}\,w(k)\left(\alpha\ket{0}+\beta\ket{1}\right)\quad \mbox{a.e. on }[-\pi,\pi],\label{eq:fourierexpand}
\end{align}
because we have
\begin{equation}
 \sum_{x\in\mathbb{Z}}\left(\frac{1}{2\pi}\int_{-\pi}^{\pi}w(\tilde{k})e^{i\tilde{k}x}d\tilde{k}\right)e^{-ikx}=w(k)
\end{equation}
almost everywhere on $[-\pi,\pi]$~\cite{JorsboeMejlbro1982,Reyna2002}.
Equation~(\ref{eq:fourierexpand}) is obtained by the Fourier series expansion.
The function $\sqrt{2\pi/W(w)}\,w(k)$ corresponds to the function $F(k)$ in Sec.~\ref{sec:limit_th}.
Note that the function $w(k)=1$ gives the QW starting from the origin with $\ket{\psi_0(0)}=\alpha\ket{0}+\beta\ket{1}$.
In the next section, we will extract the effects of this non-localized initial state in limit distributions.

\section{Limit distributions}
\label{sec:example}

Limit distribution of the QW starting from the origin with $\ket{\psi_0(0)}=\alpha\ket{0}+\beta\ket{1}$ was derived in Konno~\cite{Konno2002,Konno2005}.
If we pick $w(k)=1$, his result is obtained,
\begin{equation}
 \lim_{t\to\infty}\mathbb{E}\bigg[\bigg(\frac{X_t}{t}\biggr)^r\biggr]=\int_{-\infty}^\infty  x^r \frac{|s|}{\pi(1-x^2)\sqrt{c^2-x^2}}\biggl[1-\biggl\{|\alpha|^2-|\beta|^2+\frac{2s\Re(\alpha\overline{\beta})}{c}\biggr\}x\biggr] I_{(-|c|,|c|)}(x)\,dx.
\end{equation}
In this section, given the function $w(k)\in L^2[-\pi,\pi]$, we realize limit distributions in quantum central limit theorems from the discrete-time 2-state QW with a non-localized initial state.

\subsection{Wigner semicircle law}
If we set
\begin{equation}
 w(k)=\frac{\sin k}{1-c^2\sin^2 k},
\end{equation}
a limit theorem follows from Corollary~\ref{cor:1},
\begin{equation}
 \lim_{t\to\infty}\mathbb{E}\left[\left(\frac{X_t}{t}\right)^r\right]=\int_{-\infty}^{\infty}x^r\frac{2}{\pi c^2}\sqrt{c^2-x^2}\biggl[1-\left\{|\alpha|^2-|\beta|^2+\frac{2s\Re(\alpha\overline{\beta})}{c}\right\}x\biggr]I_{(-|c|,|c|)}(x)\,dx.\label{eq:result_case1}
\end{equation}
Note that $W(w)=\pi/|s|^3$.
When we choose $\alpha,\beta$ such that $|\alpha|^2-|\beta|^2+\frac{2s\Re(\alpha\overline{\beta})}{c}=0$ (e.g. $\alpha=1/\sqrt{2},\beta=i/\sqrt{2}$), the limit density function becomes the Wigner semicircle distribution, which is also gotten in the free central limit theorem associated to the free independence~\cite{Voiculescu1985}.

\subsection{Arcsine law}
The non-localized initial state determined by the function
\begin{equation}
 w(k)=\frac{1}{\sqrt{1-c^2\sin^2 k}}
\end{equation}
yields a convergence theorem,
\begin{equation}
 \lim_{t\to\infty}\mathbb{E}\left[\left(\frac{X_t}{t}\right)^r\right]=\int_{-\infty}^{\infty}x^r\frac{1}{\pi\sqrt{c^2-x^2}}\biggl[1-\left\{|\alpha|^2-|\beta|^2+\frac{2s\Re(\alpha\overline{\beta})}{c}\right\}x\biggr]I_{(-|c|,|c|)}(x)\,dx.\label{eq:result_case2}
\end{equation}
We should note $W(w)=2\pi/|s|$.
The density function in Eq.~(\ref{eq:result_case2}) includes the arcsine law which also appears in the monotone central limit theorem~\cite{Lu1997,Muraki1997}.

\subsection{Gaussian distribution}
Assuming
\begin{equation}
 w(k)=\sqrt{\frac{|\sin k|}{\left(1-c^2\sin^2 k\right)^{\frac{3}{2}}}}\exp\left\{-\frac{c^2\cos^2 k}{4\sigma^2\left(1-c^2\sin^2 k\right)}\right\},\label{eq:w_case3}
\end{equation}
then we have $W(w)=\frac{2\sqrt{2\pi}\sigma}{|c|s^2}\erf\left(\frac{|c|}{\sqrt{2}\sigma}\right)$ and get a limit theorem,
\begin{equation}
 \lim_{t\to\infty}\mathbb{E}\left[\left(\frac{X_t}{t}\right)^r\right]=\int_{-\infty}^{\infty}x^r\frac{\exp\left(-\frac{x^2}{2\sigma^2}\right)}{\sqrt{2\pi}\sigma\erf\left(\frac{|c|}{\sqrt{2}\sigma}\right)}\biggl[1-\left\{|\alpha|^2-|\beta|^2+\frac{2s\Re(\alpha\overline{\beta})}{c}\right\}x\biggr]I_{(-|c|,|c|)}(x)\,dx,\label{eq:result_case3}
\end{equation}
where $\sigma$ is a positive constant and the function $\erf(x)=\frac{2}{\sqrt{\pi}}\int_0^x e^{-t^2}\,dt$ denotes the Gauss error function.
This result implies Eq.~(\ref{eq:w_case3}) can lead the QW to the Gaussian distribution as $t\to\infty$.
The commuting central limit theorem associated to the commuting independence is represented by the Gaussian distribution~\cite{CushenHudson1971,GiriWaldenfels1978}.

\subsection{Uniform distribution}
In this subsection, we show that the uniform distribution can be obtained by the QW.
The uniform distribution isn't related with the central limit theorems associated to the independence in quantum probability theory.
However, from the view point of quantum computation, it is interesting for the QW to generate the uniform distribution.
Given the function
\begin{equation}
 w(k)=\sqrt{\frac{|\sin k|}{\left(1-c^2\sin^2 k\right)^\frac{3}{2}}},
\end{equation}
we get the following limit theorem,
\begin{equation}
 \lim_{t\to\infty}\mathbb{E}\left[\left(\frac{X_t}{t}\right)^r\right]=\int_{-\infty}^{\infty}x^r\frac{1}{2|c|}\biggl[1-\left\{|\alpha|^2-|\beta|^2+\frac{2s\Re(\alpha\overline{\beta})}{c}\right\}x\biggr]I_{(-|c|,|c|)}(x)\,dx.\label{eq:result_case4}
\end{equation}
Remark that $W(w)=4/s^2$.
Equation~(\ref{eq:result_case4}) proves that the QW produces the uniform distribution. 
The relation between a QW with a non-localized initial state and the uniform distribution was also discussed numerically in Valc\'{a}rcel et al.~\cite{ValcarcelRoldanRomanelli2010}.

\section{Summary}
\label{sec:summary}
\begin{figure}[h]
 \begin{center}
  \begin{minipage}{30mm}
   \begin{center}
    \includegraphics[scale=0.25]{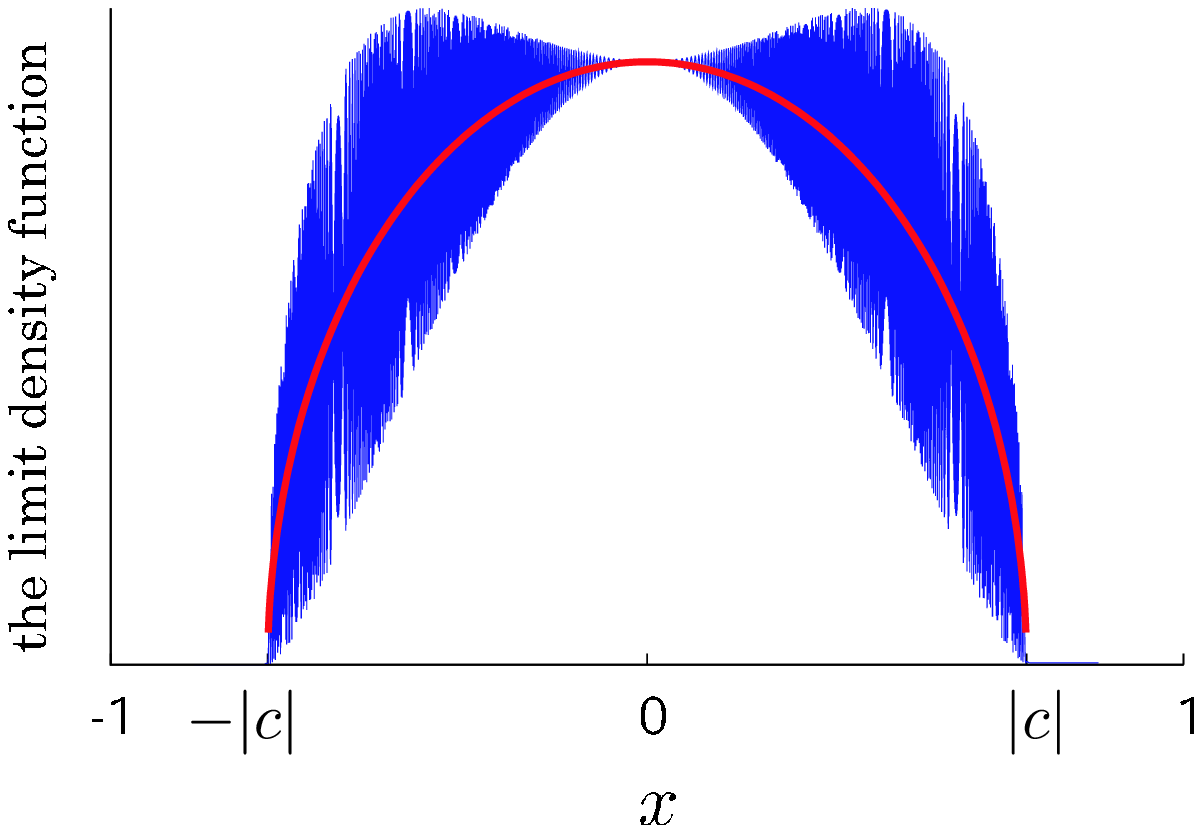}\\
    {(a) semicircle}
   \end{center}
  \end{minipage}
  \begin{minipage}{30mm}
   \begin{center}
    \includegraphics[scale=0.25]{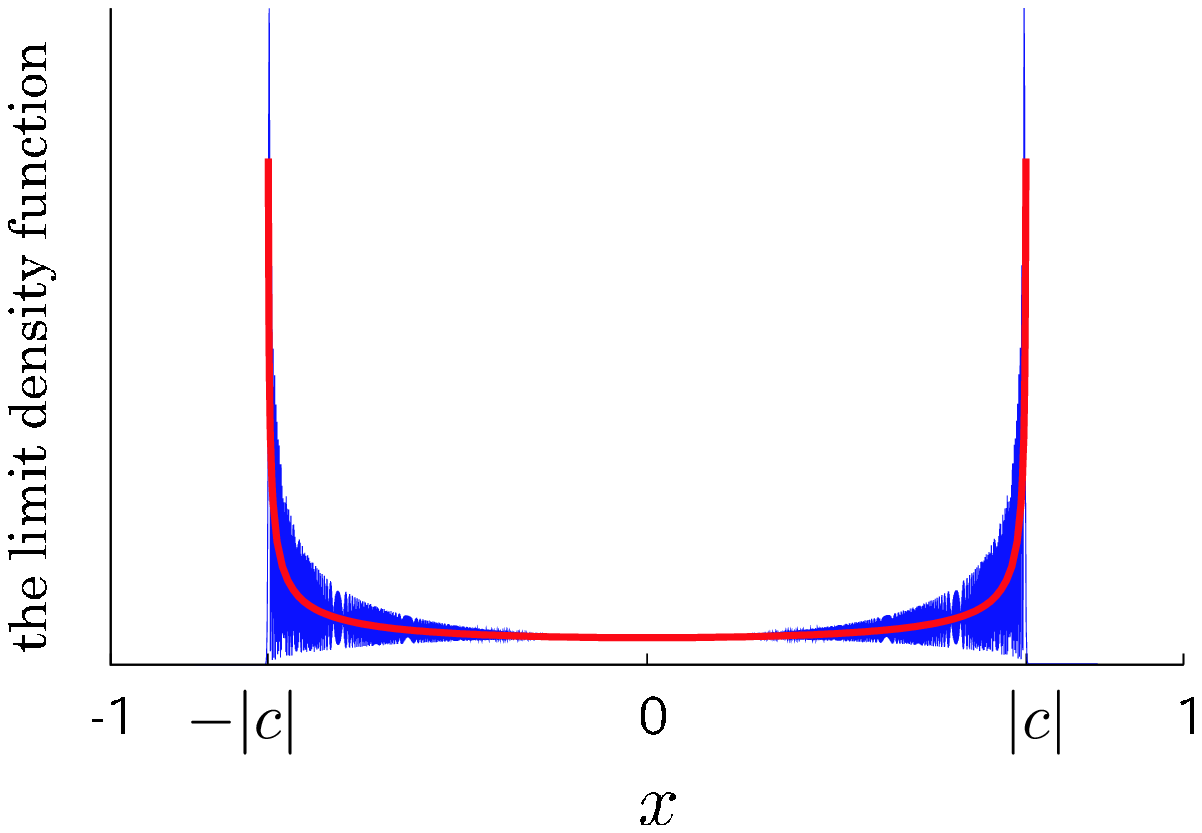}\\
    {(b) arcsine}
   \end{center}
  \end{minipage}
  \begin{minipage}{30mm}
   \begin{center}
    \includegraphics[scale=0.25]{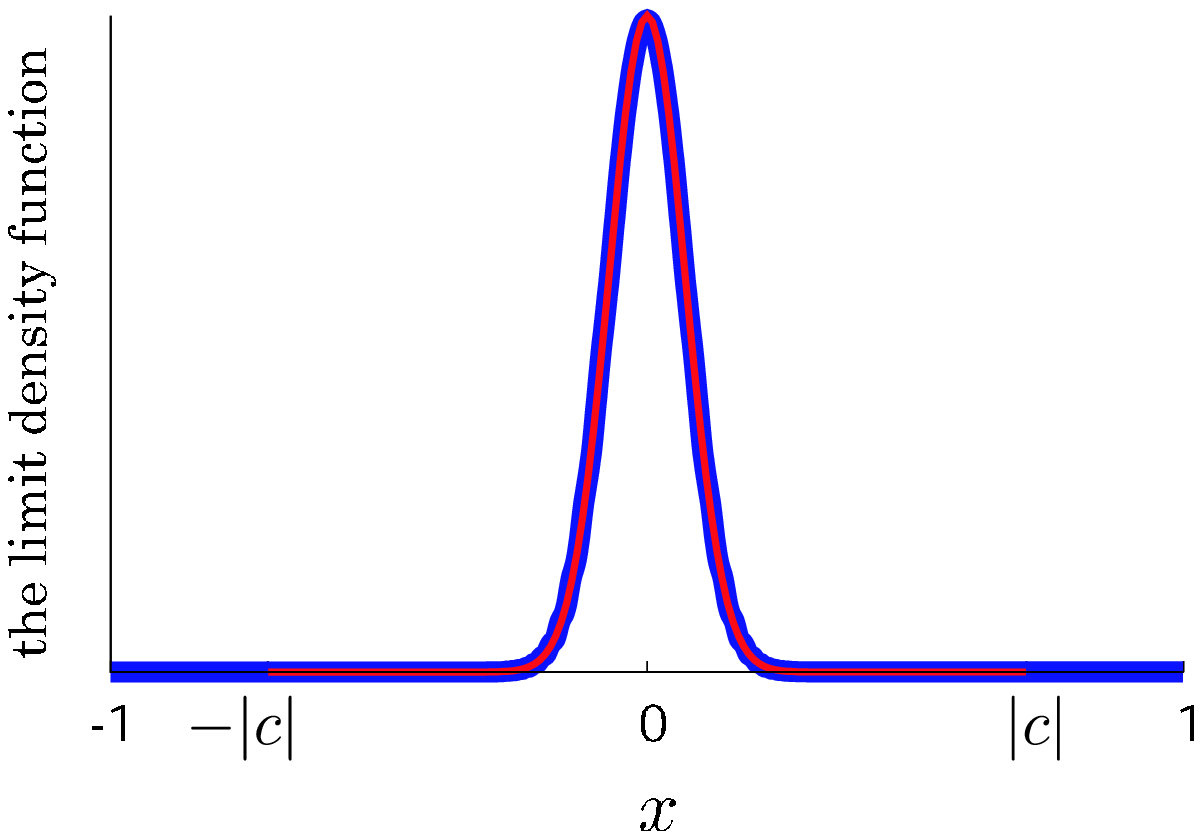}\\
    {(c) Gaussian}
   \end{center}
  \end{minipage}
  \begin{minipage}{30mm}
   \begin{center}
    \includegraphics[scale=0.25]{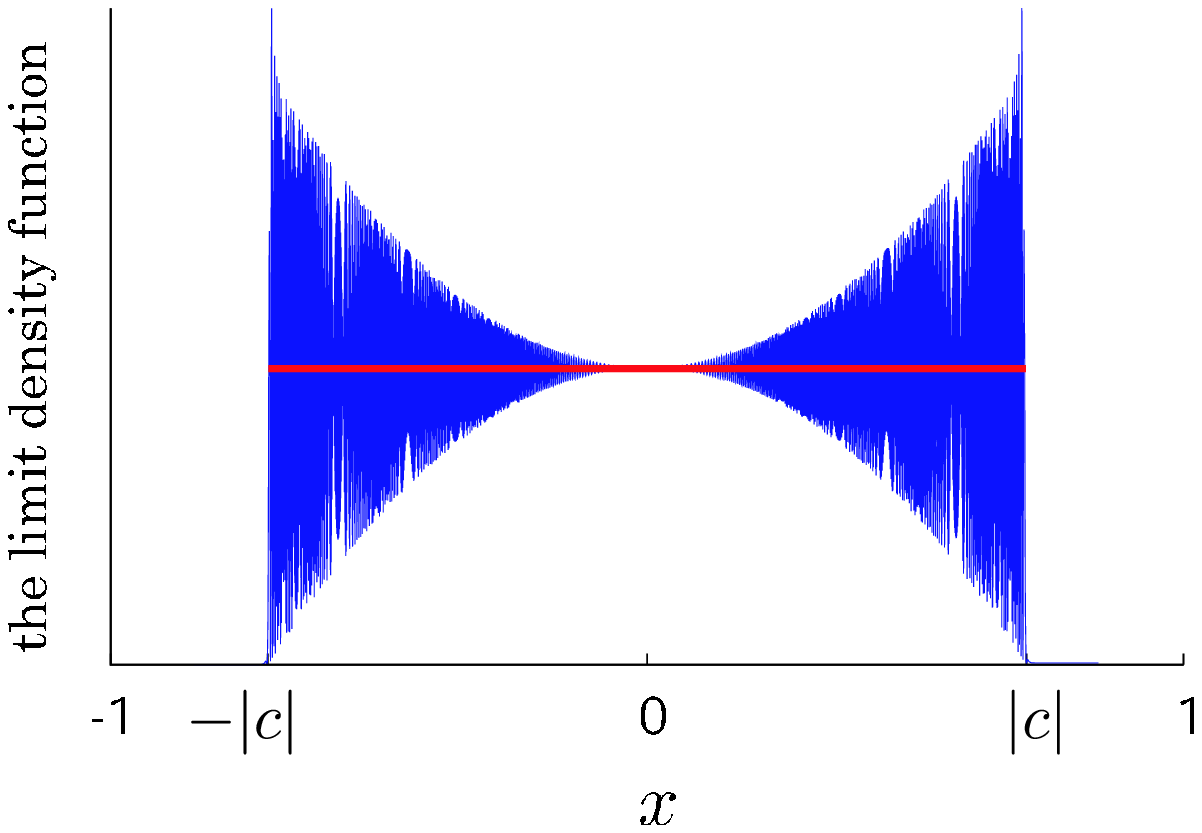}\\
    {(d) uniform}
   \end{center}
  \end{minipage}
 \end{center}
 \vspace{1mm}
 \fcaption{Comparisons between the limit density function (red line) and the probability distribution $\mathbb{P}(X_t/t=x)$ at time $t=5000$ (blue line) with $\alpha=1/\sqrt{2},\,\beta=i/\sqrt{2}$.}
 \label{fig:comparison}
\end{figure}

In this section, we argue about the conclusion and discussion for our results.
To discover a connection between QWs and quantum probability theory, we focused on a QW with a non-localized initial state.
In quantum probability theory, there are four notions of independence currently, and the quantum central limit theorems associated to them have been obtained.
Probability density functions in the central limit theorems don't agree with limit distributions of the QWs with a localized initial state.
So, we turned to limit distributions of a discrete-time 2-state QW with a non-localized initial state.
As a result, we showed that the QW can create the probability laws in the quantum central theorems.
If we take $\alpha=1/\sqrt{2}, \beta=i/\sqrt{2}$, then the density functions in Eqs.~(\ref{eq:result_case1}), (\ref{eq:result_case2}) and (\ref{eq:result_case3}) consist with those in the central limit theorems induced by the free, monotone and commuting independence, respectively.
We remark that when the QW operated by $U=\ket{0}\bra{0}-\ket{1}\bra{1}$ (i.e. $\theta=0$ case) starts from the origin, $\lim_{t\to\infty}\mathbb{P}(X_t/t\leq x)=\int_{-\infty}^x\,\frac{1}{2}(\delta_{-1}(y)+\delta_1(y))\,dy$, where $\delta_x(y)$ means the Dirac delta function.
Also, the boolean central limit theorem exhibits this probability law~\cite{SpeicherWoroudi1997}.
So, we have found that four probability laws in quantum central limit theorems can be created by the QW.
It, however, is not understood what the non-localized initial states given in this paper mean to the definition of independence in quantum probability theory.
The relation between the initial states and independence should be clear in the future. 
Moreover our limit theorem supports that we can realize some desired probability density functions with a compact support in quantum computation by controlling the function $F(k)$.
As an example of the application, we generated the uniform random valuable by the QW (see Eq.~(\ref{eq:result_case4})).   
Figure~\ref{fig:comparison} illustrates comparisons between the limit density function and the probability distribution at time $t=5000$ in the case of $\alpha=1/\sqrt{2},\,\beta=i/\sqrt{2}$.

One of the interesting future problems is to discuss convergence rate in the quantum central limit theorems and the limit theorem of the QW.
The rate of our limit theorem is $t$, while that of the quantum central limit theorems is $\sqrt{t}$.
To clarify the meaning of the difference would advance the relationship between the QWs and quantum probability theory to the next stage.

\nonumsection{Acknowledgements}
\noindent The author acknowledges support from the Meiji University Global COE Program ``Formation and Development of Mathematical Sciences Based on Modeling and Analysis''.


\end{document}